# A field-standardized application of DEA to national-scale research assessment of universities[1]

*Giovanni Abramo[a,b,\*], Tindaro Cicero[b], Ciriaco Andrea D'Angelo[b],*

[a] National Research Council of Italy

[b] Laboratory for Studies of Research and Technology Transfer
School of Engineering, Department of Management
University of Rome "Tor Vergata"

**Abstract**

The current work proposes an application of DEA methodology for measurement of technical and allocative efficiency of university research activity. The analysis is based on bibliometric data from the Italian university system for the five year period 2004-2008. Technical and allocative efficiency is measured with input being considered as a university's research staff, classified according to academic rank, and with output considered as the field-standardized impact of the research product realized by these staff. The analysis is applied to all scientific disciplines of the so-called hard sciences, and conducted at subfield level, thus at a greater level of detail than ever before achieved in national-scale research assessments.

\* **Corresponding author**, Dipartimento di Ingegneria dell'Impresa, Università degli Studi di Roma "Tor Vergata", Via del Politecnico 1, 00133 Rome - ITALY, tel/fax +39 06 72597362, giovanni.abramo@uniroma2.it

# 1. Introduction

The issue of research evaluation is attracting growing interest, involving scholars, policy makers and administrators of research institutions. A growing number of nations are carrying out periodic national research evaluation exercises which, in many cases, inform performance-based research funding. Currently, there are essentially two methodologies for national-level measurement and comparison of institutional performance: peer review and bibliometrics. Bibliometric methods have received a strong boost thanks to development of ever more sophisticated indicators and techniques of elaboration. The view of the authors is that bibliometrics will completely replace peer-review in national research evaluation exercises for the hard sciences. This trend is already seen in new evaluation exercises. The former UK Research Assessment Exercises is soon to be replaced, in 2014, by the Research Evaluation Framework, which will take an informed peer-review approach, meaning that reviewers can draw on bibliometric indicators, where appropriate, to support judgments of the quality of publications under evaluation. In Italy, the first Triennial Research Assessment (VTR, 2006) was entirely peer-review in approach, but the upcoming Quinquennial Research Evaluation exercise (expected in 2011), will allow the evaluation panels to choose between adopting only peer-review, only bibliometrics or both. The Excellence in Research for Australia initiative, launched in June 2010, provides that universities submit the entirety of their publications, and in the hard sciences these are evaluated using only bibliometric indicators. The literature justifies and supports this advancement of bibliometric practices in evaluation. For obvious reasons of costs and time it would in fact be unthinkable to utilize peer-review to evaluate the entire output of a national research system. This results in undeniable penalties concerning the performance of peer-review as compared to bibliometric method. First, it prevents any measure of productivity, the quintessential indicator of efficiency for any production system, and restricts evaluations to considering quality alone. Abramo and D'Angelo (2011) have demonstrated the superiority of bibliometrics over peer-review along other dimensions as well, namely validity, robustness, functionality, costs and time of execution.

In this paper we exploit the versatility of bibliometrics to assess the research productivity in the hard sciences of the Italian universities. Data envelopment analysis (DEA) seems a methodology that is particularly indicated for comparing efficiency of research institutions, especially given the increasing availability of quantitative indicators for input and output. In fact, there have already been signs of interest in applying DEA to measurement of research efficiency. In 1988, Ahn et al. (1988) applied DEA to 161 doctorate-granting universities in the U.S.A., to compare efficiency of public and private institutions. The study uses three types of output: undergraduate and graduate students, federal research grants and contracts. Instructional expenditures, overhead expenditures and physical investments were considered as the three inputs. The analysis did not distinguish among the various research disciplines, classifying the universities in only two groups of "with" and "without medical schools". Ten years later, Glass et al. (1998) used panel data referring to 54 publicly funded UK universities for analyzing change relative to government policy specifically designed to enhance production efficiency. Here, the DEA model consists of three outputs (one concerning research, two for teaching activities) and four inputs (identifying three labor levels and total budget). Abbott and Doucouliagos (2003) applied DEA to estimate technical and scale efficiency of 38 Australian public universities for the year 1995. They considered multiple outputs, subdivided as teaching outputs and research outcomes. Again, the organizations evaluated were simply subdivided into medical and non-medical universities. In recent years, a number of researchers in Asian nations have been quite active in conducting such studies (Xu, 2009; Yang et al., 2009; Kao et al., 2008). Meng et al. (2008) proposed a DEA model featuring hierarchical structure of input-output data and applied it to 15 research institutes of the Chinese Academy of Sciences, all active in basic research. In this study, many different outputs (publications, invited talks, awards, patents, reports, external funding, excellent leaders, graduates) are measured to produce relatively comprehensive performance profiles of these institutes. But in order to improve DEA discrimination power the output indicators are aggregated by applying the technique of analytic hierarchy process (Saaty et al, 2000). Inputs include the research staff, equipment and total research expenditure. Assessed institutes are active in "basic research", but no field-standardization has been carried out to avoid performance distortions.

There are also other works that observe individual organizational units within research institutions, such as departments. For example Tomkins and Green (1988) evaluated the cost efficiency of UK departments of accountancy for the 1984-1985 biennium, through a DEA model with four outputs (one for research and three for teaching activity) and six inputs (three for labor and three for capital). Johnes and Johnes (1993) assessed the research productivity of 36 UK departments of economics over the period 1984-88, using 32 input-output combinations. In this case the inputs express four distinct levels of departmental staff in roles that involve research activity, while the eight outputs take account of various codifications of the research results produced (articles and letters in academic journals, articles in professional journals, articles in popular journals, authored and edited books, published official reports and contributions to edited works). Beasley (1995) focused on UK Chemistry and Physics departments using a multi-output and multi-input DEA model that takes account of both teaching and research. Among the outputs, the author also inserts four dummy variables for department ratings (outstanding, above average, average or below average) as indicated by the University Grants Committee (UGC, 1986). The inputs are all of financial type. Madden et al. (1997) analyzed the effect of policy changes on the efficiency of 15 Australian economics departments from 1987 to 1991, considering both research output, teaching output and a single input for total academic staff. Finally, Bonaccorsi and Daraio (2003) applied DEA for the analysis of efficiency and comparison between 213 France INSERM institutes and 27 institutes of the Italian National Research Council for the year 1997. In this case the focus was only on research and the authors proposed a new input variable, the Geographical Agglomeration Index, which measures the geographic concentration of these institutes, in order to account for the influence of proximity and possible advantages of location. However the analysis does not feature field-standardization and considers organizations from the fields of medicine, biology and biomedical molecular biology equally.

The particular flexibility of DEA for measurement of production efficiency in multi-output, multi-input processes has led many scholars to concentrate their efforts on assembling as much data as possible on potential outputs on the one hand and production factors on the other, while at the same time ignoring other fundamental aspects such as scientific uniformity of the decision making unit (DMU) evaluated. We refer in particular to the studies conducted at the level of entire organizations, which typically do not consider appropriate field-standardizations. The literature provides ample demonstration of the varying intensity of publication and citation among scientific sectors, and the distortion in research efficiency measures that result when this is not considered (Abramo et al., 2008a). The studies conducted at the level of single departments are only partially exempt from this problem, and their value is limited by the difficulty of applying such studies across all disciplines that concern policy-makers or administrators of large research institutions. To the authors' knowledge, the only attempt to apply DEA on a national scale, at the level of macro disciplines, is that conducted by Abramo et al. (2008b) on all Italian universities active in the hard sciences. However, in light of recent advancements in the field of bibliometric techniques, that study appears subject to criticism concerning i) level of detail of the analysis (macro disciplines); ii) types of output indicator used (publications, co-authorship and journal impact factor). These are issues that the current work is intended to overcome by: i) carrying out the elaborations at the micro level of single scientific subfields; ii) considering a single measure of output concerning knowledge advancement generated by the university, and proxied by field-standardized citations received by publications.

The analysis proposed in the present work is based on the observation of the entire Italian university system over the five year period 2004-2008, meaning examination of 78 universities and 130 scientific subfields. The great breadth of the field of observation imposes some simplifying hypothesis in the formulation of inputs and outputs for the DEA model, which will be shown to be completely acceptable for the Italian context. However, with respect to previous literature, the intention of the study is to propose a national-scale comparative measurement of university technical and allocative efficiency, free of any distortion due to lack of field-standardization. Given that the objective of analysis is to support stimulus interventions for greater efficiency, then such measures at a detailed level are certainly opportune, because they provide indicative feedback concerning the individual actors in the system (Abramo and D'Angelo, 2010). By measuring the efficiency only of research organizations, we are not suggesting that other performance indicators should be disregarded. It is up to governments to decide which indicators to use and their relative weight.

The next section of this work describes certain key characteristics of the Italian university system and then presents the DEA model used, with the input and output variables and the relative data sources. The third and fourth sections discuss the results from application of the DEA, through presentation of examples: Section 3 presents the results for evaluation of technical and allocative efficiency of universities active in a specific scientific subfield, while Section 4 focuses on the evaluation of efficiency for all scientific subfields of a university. The closing section offers the authors considerations on possible application and enhancement of the proposed model for various decision-making levels, and of possible further developments.

## 2. Methodology and data

Research activity is a production process in which the inputs consist of human, tangible (scientific instruments, materials, etc.) and intangible (accumulated knowledge, social networks, etc.) resources, and where outputs have a complex character of both tangible nature (publications, patents, conference presentations, databases, protocols, etc.) and intangible nature (tacit knowledge, consulting activity, etc.). The new-knowledge production function has therefore a multi-input and multi-output character. This in turn creates a multi-faceted problem when it comes to measuring efficiency, and requires scholars to make precise choices in methodology. These choices must obviously relate to the detailed context for the analysis, a context which we describe in the following section.

### 2.1 The Italian academic system

In Italy, the Ministry of Education, Universities and Research (MIUR) recognizes a total of 95 universities as having the authority to issue legally-recognized degrees. With only rare exceptions these are public universities largely financed through non-competitive allocation. Up to 2009, the core government funding was input oriented, i.e. distributed to universities in a manner intended to equally satisfy the needs and resources of each and all, in function of their size and activities. The share of this core funding relative to total university income is now being reduced, descending from 61.5% in 2001 to 55.5% in 2007 (MIUR, 2009). It was only following the first national research evaluation exercise (VTR), conducted between 2004 and 2006, that a minimal share, equivalent to 7% of MIUR financing[2], was attributed in function of the assessment of research and teaching quality. Further financing from the MIUR for research projects on a competitive basis[3] represents an additional 9% of income. Other public and private financing for research projects, obtained on a competitive basis, adds a further 17% of total income. The government imposes a price-cap for tuition fees. They vary from student to student according to family income, however they are very low and provide only about 12.5% of total university income. Income deriving from technological transfer is negligible, given the very limited practice of Italian universities to carry out patenting and licensing (Abramo, 2006). Donations to universities are not a feature of the Italian tradition and are also negligible. All new personnel enter the university system through public examinations, and career advancement also requires such public examinations. Salaries are regulated at the nationally centralized level and are calculated according to role (administrative, technical, or professorial), rank within role (for example: assistant, associate or full professor), and seniority. No part of the salary for professors is related to merit: wages are increased annually according to parameters set by government. All professors are contractually obligated to carry out research, thus all universities are research universities: "teaching-only" universities do not exist. Each research staff member is classified in one and only one scientific disciplinary sector (SDS), 370 in all. SDSs are grouped in 14 University Disciplinary Areas (UDAs).

---

[2] Since MIUR financing composes 55.5% of the total, the share distributed on the basis of the VTR represents 3.9% of total income.

[3] In the Italian university sphere there is great skepticism that public-sector financing of research is truly determined on a merit basis, due to a perceived history of strongly-rooted favoritism and of apparent ineffectiveness in project evaluation procedures.

**2.2 The new-knowledge production function**

The analysis of research efficiency will be conducted at the level of the SDS: a DMU thus consists of the research staff in a university SDS and for each SDS the DEA analysis is based on the input and output data for all national universities that have research staff active in that SDS. The simplified new-knowledge production function that we propose for study of the Italian case considers human resources as the sole input. We exclude capital on the basis of the information presented above, specifically that in the Italian university system, the large part of financial resources is equally allocated by government to satisfy the needs of each university in function of its size. The potential greater availability of funds per staff unit in a DMU is thus due to its capacity to acquire such funds on a competitive basis. Greater output deriving from greater availability of funds is thus the result of merit and not of any other comparative advantages. One can thus hold that, other than the quality of their research staff, it is the capacity to attract such funding and its efficient use that explains the difference in productivity among DMUs. Ideally, one should not ignore the intangible factor of possible advantages in location and economic rents for universities situated in areas of high intensity for private research, where geographic proximity effect (Jaffe, 1989; Lang, 2005; Coronado and Acosta, 2005) would permit greater opportunities for private financing and stimuli. However relevant data at the level of these DMUs is unavailable, preventing the authors from taking this factor into consideration.

The universities have essentially three missions: education, advancement of scientific-technological knowledge through research, and support of industrial competitiveness through technology transfer. Even if these three duties could be considered mutually reinforcing, in reality they compete directly for the available time of each single scientist. It is thus possible that the time dedicated to research by the DMU staff is not uniform among all, but data on the facts of this is not available. The evaluation of a university would thus ideally be conducted for the three dimensions that characterize its mission. Lesser efficiency in research activity of a DMU could in reality imply greater dedication to other activities. However, this current study will measure production efficiency only in the universities' research activity.

Limiting the field of analysis to the hard sciences, we can choose a pure bibliometric approach and consider the citations of scientific publications in international journals as proxy of impact on advancement of knowledge generated by research activity. Despite the inherent limits of this proxy, it is seen as the most reliable and representative for large-scale comparative analyses (Moed et al., 2004). Further, the choice of excluding other recognized outputs (authored and edited books, reports, patents, prototypes, etc.) also has clear empirical support, given that in the first national research evaluation for Italy (VTR, 2006), journal articles alone represented 95% of total research outputs submitted by universities for assessment in the hard sciences (Abramo et al., 2009). Certain other codifications of research output are both difficult to census and would offer information of very slight marginal utility. For example, patent applications might provide a useful complement, but in Italy the number of patents filed by all universities is notoriously low (less than 1,500 between 2001 and 2007). Since the introduction of the academic privilege in 2001[4] it appears that there have indeed been a higher number of patents filed by university faculty members, but relevant, meaningful data are not available and analysis of the issue thus remains infeasible. Further, the literature demonstrates that there is a high correlation between intensity of publication and patenting by researchers (Adams and Griliches, 1998; Lach and Schankerman, 2003; Van Looy et al., 2006). And finally, patents are often followed by publications that describe their content in the scientific arena, so the analysis of publications alone actually avoids a potential double counting. In any case, patenting and licensing should be examined in evaluation of the separate dimension of technology transfer activity.

**2.3 DEA specifications and data source**

The DEA model is based on: i) three input variables concerning the research staff of a university, distinguished according to three faculty ranks: assistant, associate and full professors; ii) one output variable, concerning the impact on knowledge advancement generated by the university and proxied by the bibliometric indicator "Scientific Strength" (*SS*), i.e. field-standardized citations received by publications authored by the research staff of the university. Here:

$$SS = \sum_{i=1}^{n} \bar{C}_i \cdot f_i$$

Where:
$\bar{C}_i$ = standardized citations of publication *i* by the university. Citations of each publication are standardized dividing them by the median[5] of citations[6] of all Italian publications of the same year and WoS subject category[7].
$f_i$ = fractional count of publication *i*, i.e. the ratio of the number of co-authors of the university to the total number of co-authors. For life sciences, different weights have been given to each co-author of the university according to his/her position in the list and the character of the co-authorship (intra-mural or extra-mural)[8].

---

[4] Under the Italian Law 383/2001, intellectual property rights on public employees' inventions are granted to the employees themselves.
[5] The decision to standardize citations with respect to the median number is justified by the fact that the distribution of citations is highly skewed (Lundberg, 2007). A possible alternative would be to standardize to the world average, as frequently observed in the literature.
[6] Observed as of 30/06/2009.
[7] When a publication falls in two or more subject categories the average of the medians is used.
[8] If first and last authors belong to the same university, 40% of citations are attributed to each of them; the remaining 20% are divided among all other authors. If the first two and last two authors belong to different

$n$ = number of publications by the university.

Table 1 summarizes the input and output variables of the DEA model.

| Variable | Type | Acronym |
|---|---|---|
| Staff-years of full professors | Input | FP |
| Staff-years of associate professors | Input | AP |
| Staff-years of assistant professors | Input | RF |
| Scientific Strength | Output | SS |

*Table 1: Variables for the input-oriented DEA model*

Italian universities are not homogenous in terms of fields of research investigation. Standardizing citations of each publication is not sufficient then to avoid distortions in productivity assessment, because of the varying intensity of publications in the various disciplines and fields[9]. To account for that, we carried out the analysis at SDS level[10]: the Italian university system provides for 205 SDSs[11], grouped into nine UDAs[12], that in turn represent the entirety of the "hard sciences".

The data source is the Italian Observatory of Public Research (ORP)[13], a database developed and maintained by the authors, derived under license from the Thomson Reuters Web of Science (WoS). Beginning from the WoS data and applying a complex algorithm for disambiguation of the true identity of the authors and reconciliation of their institutional affiliations, each publication (article, review or conference proceeding) is attributed to the university scientist that produced it (D'Angelo et al., 2010) For the current analysis, the SDS of a university is the DMU and its output is equal to the SS calculated for the scientific production of all scientists that belong to that SDS. The input is represented by the research staff at the university who belong to the SDS/DMU under examination. The chosen level of analysis (SDS) permits due account for the varying intensity of publication and citation in the different areas, thus overcoming the evident limits of all previous studies found in the literature. For a more significant and robust analysis the field of observation is limited to SDSs where:

- at least 50% of Italian universities' scientists produced at least one ORP publication in the period 2004-2008 (significance of scientific publication as proxy of research output);
- there are at least 24 universities active at the national level (robustness of the DEA application).

For the period under consideration, the dataset is thus composed of 130 SDSs with an average research staff in the period of observation of 32,028 individuals, including

---

universities, 30% of citations are attributed to first and last authors; 15% of citations are attributed to second and last author but one; the remaining 10%.are divided among all others.

[9] The average number of yearly publications of a physicist are 2.3 times as many as those of a mathematician.
[10] Although members of the same SDS may publish in different WoS subject categories, their publication rate is not affected so much by their field of research. Differences in SS reflect differences in productivity.
[11] The complete list is accessible on http://attiministeriali.miur.it/UserFiles/115.htm
[12] Mathematics and computer sciences; physics; chemistry; earth sciences; biology; medicine; agricultural and veterinary sciences; civil engineering; industrial and information engineering.
[13] www.orp.researchvalue.it.

assistant, associate and full professors, distributed among 78 universities (Table 2).

| UDA | N. of SDSs | Universities | Professors | | | Total |
|---|---|---|---|---|---|---|
| | | | Assistant | Associate | Full | |
| Mathematics and computer sciences | 8 | 64 | 970 | 1,092 | 1,083 | 3,145 |
| Physics | 6 | 61 | 677 | 928 | 874 | 2,479 |
| Chemistry | 9 | 59 | 936 | 1,092 | 1,023 | 3,050 |
| Earth sciences | 10 | 48 | 349 | 442 | 404 | 1,195 |
| Biology | 19 | 66 | 1,874 | 1,623 | 1,663 | 5,160 |
| Medicine | 43 | 56 | 4,658 | 3,267 | 2,791 | 10,716 |
| Agricultural and veterinary sciences | 28 | 40 | 340 | 283 | 378 | 1,001 |
| Civil engineering and architecture | 7 | 49 | 388 | 434 | 480 | 1,302 |
| Industrial and information engineering | 21 | 66 | 1,119 | 1,252 | 1,609 | 3,980 |
| Total | 130 | 78 | 11,312 | 10,411 | 10,305 | 32,028 |

*Table 2: Number of universities, UDAs, SDSs, and professors per academic rank of the dataset*

The elaborations were conducted using the *Data Envelopment Analysis Program* (Coelli, 1996). This tool allows calculation of cost efficiency (CE) scores, which equal the radial distance of each DMU from the efficient frontier, with the following hypotheses:

i. <u>Constant returns to scale</u>: both literature analysis and tests conducted by the current authors lead to reject the hypothesis that returns vary with the scale of the DMU;
ii. <u>Input orientation</u>: the score represents the maximum equi-proportional decrease in all inputs (outputs remaining equal). This model reflects the management objective of reducing production factors while maintaining equal output. A score value of 1 indicates fully efficient DMUs.

The cost efficiency scores are in turn given by the product of two different factors: technical (TE) and allocative (AE) efficiency[14]. Compared to classic input-output techniques that give a single efficiency score, one of the advantages of DEA is that it provides an understanding of the weight of the two dimensions of technical and allocative efficiency in determining overall economic efficiency. For this, each production factor is weighted according to a vector of coefficients (56.650; 79.700; 111.700) that reflect that average cost (in k€) of the three inputs, meaning the cost of research staff for the three Italian academic ranks considered over the period under observation[15].

## 3. Evaluation of technical and allocative efficiency of the SDSs

The calculation of DEA efficiency scores was carried out for each of the 130 SDSs examined. In this section we present the results of the analyses through the example of the Pharmaceutical chemistry SDS (CHIM/08). Such analyses signify a valid assist, penetrating to a very detailed level, for national evaluation exercises designed to support performance-based research funding systems.

---

[14] Technical efficiency represents the ability of a DMU to maximize output subject to a given combination of production factors. Allocative efficiency represents the ability to minimize the costs of production factors to produce a given output.

[15] Data concerning salary costs were obtained from the DALIA database, maintained by the Italian Ministry of Education, Universities and Research (https://dalia.cineca.it/php4/inizio_access_cnvsu.php, last accessed on April 27, 2011).

Table 3 presents the values for input, output, and efficiency scores for every university active in this SDS. Of the 28 total DMU, six show maximum technical efficiency value (TE = 1.000), but only one of these (Univ_1) also shows a maximum value for allocative efficiency. For cost efficiency, Univ_2 and Univ_3 place very near the efficiency frontier, with scores of 0.948 and 0.945. For Univ_4 to Univ_11 the cost efficiency scores are greater than 0.6: three of these cases, Univ_5, Univ_7 and Univ_9, reach maximum technical efficiency. From Univ_12 down, all coefficients of cost efficiency are less than 0.5, meaning far from the efficiency frontier. Many of the universities show high values of allocative efficiency, but low values of technical efficiency. In addition to distinguishing between types of efficiency, the accuracy level of DEA as compared to simple output/input measures, can be further appreciated by noting that the university ranking by SS/Staff years would be different[16].

| University | SS | Staff years | | | Cost DEA* | | |
|---|---|---|---|---|---|---|---|
| | | Full | Associate | Assistant | TE | AE | CE |
| Ferrara | 64.026 | 25 | 18 | 20 | 1.000 | 1.000 | 1.000 |
| Piemonte Orient. Avogadro | 24.970 | 5 | 10 | 15 | 1.000 | 0.948 | 0.948 |
| Bologna | 151.659 | 47 | 65 | 53 | 1.000 | 0.945 | 0.945 |
| Siena | 57.508 | 25 | 24 | 19 | 0.918 | 0.907 | 0.833 |
| Pavia | 40.293 | 10 | 29 | 17 | 1.000 | 0.768 | 0.768 |
| Perugia | 42.115 | 17 | 29 | 19 | 0.772 | 0.864 | 0.667 |
| Milano | 77.785 | 50 | 20 | 46 | 1.000 | 0.666 | 0.666 |
| Roma "La Sapienza" | 80.585 | 22 | 69 | 41 | 0.882 | 0.744 | 0.656 |
| Messina | 41.627 | 24 | 30 | 8 | 1.000 | 0.631 | 0.631 |
| Padova | 62.036 | 40 | 45 | 22 | 0.729 | 0.766 | 0.558 |
| Urbino "Carlo Bo" | 22.055 | 5 | 21 | 23 | 0.883 | 0.591 | 0.522 |
| Napoli "Federico II" | 53.332 | 29 | 43 | 53 | 0.484 | 0.955 | 0.462 |
| Parma | 29.019 | 20 | 31 | 24 | 0.437 | 0.918 | 0.401 |
| Trieste | 21.015 | 13 | 27 | 17 | 0.465 | 0.829 | 0.385 |
| Modena e Reggio Emilia | 23.199 | 9 | 35 | 23 | 0.553 | 0.689 | 0.381 |
| Firenze | 46.893 | 34 | 50 | 45 | 0.404 | 0.941 | 0.380 |
| Cagliari | 16.012 | 10 | 15 | 22 | 0.396 | 0.951 | 0.377 |
| Pisa | 38.584 | 32 | 39 | 35 | 0.389 | 0.957 | 0.373 |
| Salerno | 11.998 | 10 | 15 | 17 | 0.322 | 0.951 | 0.307 |
| Torino | 27.728 | 20 | 40 | 39 | 0.341 | 0.893 | 0.304 |
| Camerino | 23.703 | 40 | 27 | 11 | 0.502 | 0.546 | 0.274 |
| Bari | 43.205 | 42 | 67 | 62 | 0.288 | 0.928 | 0.267 |
| Sassari | 18.698 | 19 | 25 | 43 | 0.266 | 0.897 | 0.239 |
| Calabria | 4.180 | 10 | 5 | 4 | 0.303 | 0.663 | 0.201 |
| Catania | 16.047 | 22 | 50 | 25 | 0.225 | 0.759 | 0.171 |
| Genova | 16.147 | 30 | 42 | 22 | 0.217 | 0.786 | 0.170 |
| Palermo | 19.871 | 40 | 40 | 48 | 0.164 | 0.979 | 0.160 |
| Gabriele D' Annunzio | 6.323 | 13 | 5 | 27 | 0.325 | 0.482 | 0.157 |

*Table 3: Efficiency scores for the 28 universities active in the Pharmaceutical chemistry SDS*

---

[16] For example, Piemonte Orientale Avogadro's SS/Staff years = 0.832 is lower than Bologna's (0.919) and Siena's (0.846). According to the SS/Staff years index, University of Piemonte Orientale Avogadro would shift from second to fourth position.

*\* TE: Technical Efficiency; AE: Allocative Efficiency; CE: Cost Efficiency*

The frequency distributions for efficiency scores obtained in this SDS are shown in Figures 1-3. The distribution for the technical efficiency scores is bimodal, with a strong concentration in the second (0.20-0.40) and last (0.80-1) intervals, while the distribution for allocative efficiency scores is markedly asymmetric to the left, with an extremely high median value (0.878). Finally, the cost efficiency scores distribution is asymmetric to the right. The modal quintile is 0.20-0.40 and the median value is relatively low (0.383).

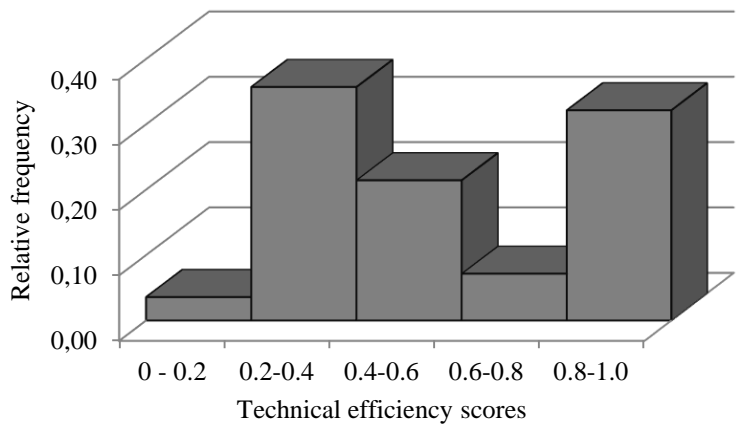

*Figure 1: Distribution of technical efficiency scores for the 28 universities active in Pharmaceutical chemistry*

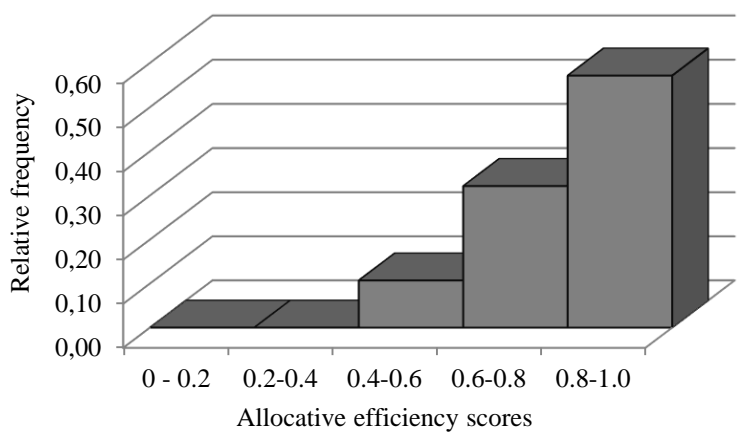

*Figure 2: Distribution of allocative efficiency scores for the 28 universities active in Pharmaceutical chemistry*

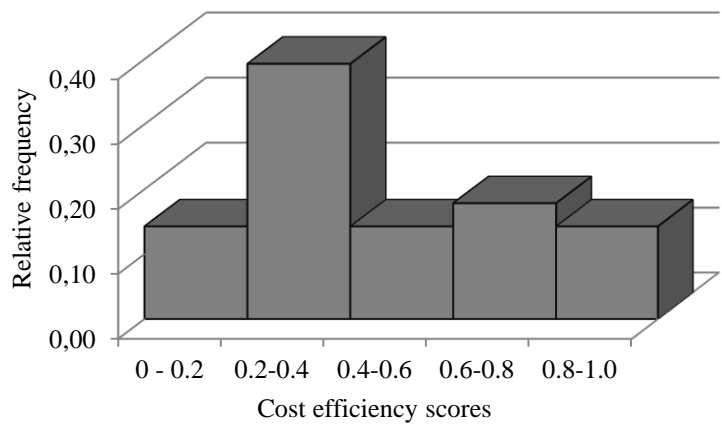

*Figure 3: Distribution of cost efficiency scores for the 28 universities active in Pharmaceutical chemistry*

In Figure 4 we relate cost efficiency scores to university size[17], with the x-axis showing university ID and y-axis indicating score. The sphere dimension represents the research staff size of the SDS at each university. The university SDS with maximum efficiency score is one that is medium-small size, with 63 staff years of full, associate and assistant professors over the five years. The largest universities show highly variable efficiency levels, and there is no clear link between size and efficiency level. For example, Univ_22, which has the largest staff in the nation (171 staff years, with 47 full, 65 associate and 53 assistant professors) shows a low level of cost efficiency (0.267), due to its low technical efficiency, while Univ_3, which is second in size (165 staff years over the 5-year period), shows a very high value for cost efficiency (0.945).

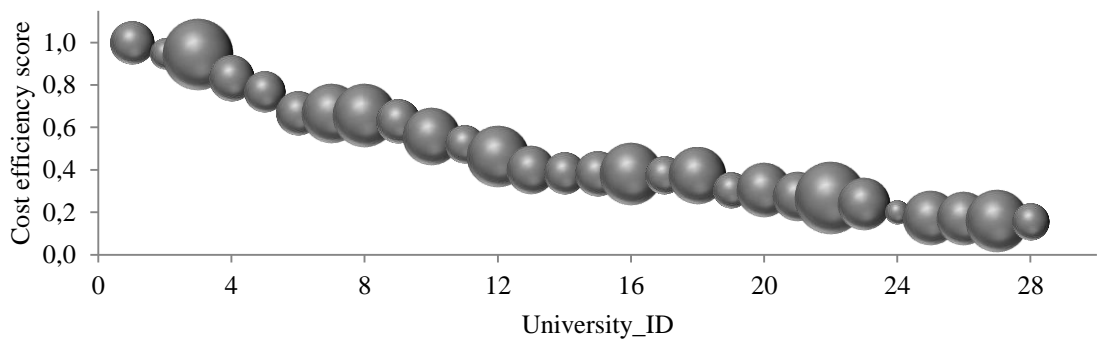

*Figure 4: Cost efficiency scores in relation to SDS size for the universities active in Pharmaceutical chemistry*

In Figure 5 we present a data matrix of the for the 28 universities active in the Pharmaceutical chemistry SDS, as a means of classifying the universities on the basis of the combination of their scores for allocative and technical efficiency.

---

[17] We assumed constant returns to scale in the DEA application. In the example we test if the assumption is confirmed, at least for the specific case of Pharmaceutical chemistry.

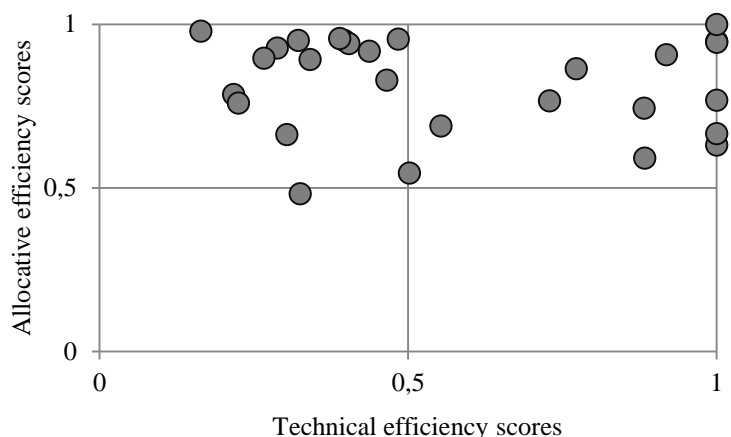
*Figure 5: Efficiency matrix of the 28 universities active in Pharmaceutical chemistry.*

Only one university presents low values for both efficiency measures. The remaining universities are almost equally divided between the two upper quadrants, with one more to the upper left (14 versus 13 to upper right), which represents high scores in allocative efficiency and low ones in technical efficiency. In general, the SDS analyzed is thus characterized by substantial uniformity in distribution of allocative efficiency compared to a significant differentiation for technical efficiency.

## 4. Efficiency evaluation of the SDSs of a university

The preceding section presented the results of an analysis for all Italian universities active in a particular SDS, taking the view of a policy-maker conducting a national research assessment and interested in allocation of resources in function of merit. In this section we take the perspective of an administrator for an individual university who wishes to evaluate its SDSs and identify strong and weak points by comparing with efficiency performances on a national scale. As example, we take an institution ("Univ_Y") with a large research staff and we analyze the efficiency scores for all the SDSs of a single UDA.

Table 4 presents the scores for technical, allocative and cost efficiency for the 19 SDSs of the Biology UDA at an institution which we call "Univ_Y". For this UDA there is a sum total of 281 labor units, with a slight prevalence of assistant professors (107) compared to full (97) and associate professors (77), and the total value for the labor production factor over the five years is just under 115 million euro. In order to compare the performance scores of the SDSs, independent of their individual characteristics, calculations were completed to rank them by percentile according to the entire distribution of scores for their relative national SDS. The only fully cost-efficient SDS is BIO/15 – Pharmaceutical biology. However, none of the SDSs show a nil CE score: the minimum value (0.007) is for BIO/17 – Histology. There are various SDSs with contrasting performances: in BIO/03 – Environmental biology, the technical efficiency is almost at the maximum (0.946), while allocative efficiency is quite low (0.376). The reverse occurs in BIO/16 – Human anatomy, where technical efficiency (0.130) is among the bottom in national rankings (42$^{nd}$ out of 47) while allocative efficiency (0.959) is among the top (7th out of 47).

Finally, we aggregate the scores for the single SDSs to arrive at an average score for the UDA. The aggregation is carried out with weighting of scores on the basis of SDS's size, per their costs of research staff, indicated in Column 2 of Table 4. The final row of the table gives the average efficiency values for the UDA and the UDA's national percentile ranking. For Biology, this university shows an overall low technical efficiency (0.239), corresponding to 28th percentile in the national rankings. The allocative efficiency is much better (0.635, corresponding to 75th national percentile), but cost efficiency remains very low (0.160), meaning that the university places under the national average, in the 42nd percentile. It is interesting and potentially useful to note that this result is primarily due to the very low performance of BIO/10. This is the largest SDS, but it is where the university sees scores for technical efficiency (0.092) and allocative efficiency (0.544) that are decidedly low. Further, the overall score for the observed UDA does not reflect the performance levels achieved by many of its SDSs, even though these may be SDSs of small size. Such results confirm the importance of conducting analysis at the most detailed possible levels of aggregation.

| SDS | Cost of research staff (k €) | Technical effic. Score | R%* | Allocative effic. Score | R% | Cost effic. Score | R% |
|---|---|---|---|---|---|---|---|
| BIO/01 | 4,4730 | 0.592 | 78 | 0.592 | 58 | 0.351 | 86 |
| BIO/02 | 3,773.50 | 0.148 | 59 | 0.486 | 71 | 0.072 | 65 |
| BIO/03 | 2,435.75 | 0.946 | 85 | 0.376 | 51 | 0.355 | 90 |
| BIO/04 | 3,880.75 | 0.079 | 18 | 0.681 | 55 | 0.054 | 32 |
| BIO/05 | 5,548.10 | 0.064 | 12 | 0.705 | 54 | 0.045 | 15 |
| BIO/06 | 13,995.50 | 0.157 | 28 | 0.805 | 72 | 0.126 | 39 |
| BIO/07 | 2,136.70 | 0.258 | 62 | 0.339 | 14 | 0.087 | 36 |
| BIO/08 | 558.50 | 0.949 | 78 | 0.680 | 78 | 0.645 | 87 |
| BIO/09 | 11,579.55 | 0.105 | 17 | 0.464 | 83 | 0.049 | 31 |
| BIO/10 | 24,884.55 | 0.092 | 51 | 0.544 | 53 | 0.050 | 58 |
| BIO/11 | 7,202.65 | 0.146 | 33 | 0.706 | 84 | 0.103 | 42 |
| BIO/12 | 4,846.35 | 0.188 | 51 | 0.506 | 78 | 0.095 | 68 |
| BIO/13 | 5,394.30 | 0.245 | 38 | 0.696 | 64 | 0.170 | 47 |
| BIO/14 | 8,859.70 | 0.676 | 78 | 0.77 | 41 | 0.521 | 83 |
| BIO/15 | 2,697.15 | 1.000 | 100 | 1.000 | 100 | 1.000 | 100 |
| BIO/16 | 2,701.95 | 0.130 | 11 | 0.959 | 87 | 0.125 | 24 |
| BIO/17 | 1,351.95 | 0.013 | 11 | 0.526 | 47 | 0.007 | 14 |
| BIO/18 | 6,575.35 | 0.150 | 24 | 0.680 | 70 | 0.102 | 24 |
| BIO/19 | 2,023.40 | 0.565 | 56 | 0.547 | 47 | 0.309 | 61 |
| *Total/average* | 114,918.70 | 0.239 | 28 | 0.635 | 75 | 0.160 | 42 |

***Table 4: DEA and percentile (R%) scores for the SDSs of the Biology UDA at Univ_Y***
*\* Percentile rank 0 – 100, with 100 as best and 0 as worst*

The logic just illustrated for evaluation of the SDSs of a UDA can be extended to all sectors where a university is active, obtaining an efficiency score (thus a national rank) for the entire university. For this, we offer the example of a small university (total input equal to 551 staff years in the five years observed), which we call "Univ_X". Table 5 presents the DEA results for the 23 SDSs in which the university is active. The table shows that the university reaches maximum efficiency in the Agronomy and herbaceous agriculture (AGR/02) SDS. In the SDSs GEO/02 – Stratigraphic geology and GEO/05 – Applied geology, in spite of cost efficiency scores that do not reach 0.500, the university places among the top 10% (percentile rank greater than 90) in national comparison with other universities. In particular, in GEO/05, Univ_X shows the maximum score for allocative efficiency; while the score for technical efficiency is lower (0.466), but still achieves 87$^{th}$ national percentile ranking.

In general we see high variability for cost efficiency among the various sectors within the university. There are three SDSs (BIO/03, BIO/05 and GEO/07) with nil output, and thus nil efficiency. Other than these three cases, the cost efficiency for this university's SDSs does not descend below the score of 0.030, seen in Biochemistry (BIO/10).

Aggregating the scores of the 23 SDSs, with weighting for the cost of research staff per Column 2 of Table 5, provides the overall average scores seen in the last row of the table. The score for technical efficiency (0.352) puts the university in the 51$^{st}$ percentile rank at national level. If input cost is also considered then the overall performance increases slightly: the university places in 52$^{nd}$ national percentile for cost efficiency (0.228).

|  | Cost of | TE | | AE | | CE | |
| --- | --- | --- | --- | --- | --- | --- | --- |
| SDS | research staff (k €) | Score | R%* | Score | R% | Score | R% |
| AGR/02 | 398.500 | 1.000 | 100 | 1 | 100 | 1 | 100 |
| BIO/03 | 957.000 | 0.000 | 0 | 0.000 | 0 | 0.000 | 0 |
| BIO/05 | 111.700 | 0.000 | 0 | 0.000 | 0 | 0.000 | 0 |
| BIO/07 | 3,826.050 | 0.251 | 60 | 0.845 | 83 | 0.212 | 71 |
| BIO/10 | 1,355.500 | 0.062 | 36 | 0.480 | 29 | 0.030 | 31 |
| BIO/19 | 1,117.000 | 0.867 | 81 | 0.167 | 11 | 0.145 | 25 |
| CHIM/01 | 5,281.000 | 0.323 | 38 | 0.832 | 93 | 0.268 | 69 |
| CHIM/02 | 4,129.200 | 0.084 | 61 | 0.426 | 10 | 0.036 | 29 |
| CHIM/03 | 7,065.550 | 0.168 | 33 | 0.639 | 27 | 0.107 | 24 |
| CHIM/06 | 3,722.700 | 0.385 | 31 | 0.884 | 90 | 0.341 | 46 |
| CHIM/12 | 2,871.000 | 0.236 | 54 | 0.603 | 68 | 0.142 | 79 |
| FIS/01 | 1,844.250 | 0.701 | 71 | 0.762 | 44 | 0.534 | 83 |
| FIS/03 | 681.750 | 0.117 | 7 | 1.000 | 100 | 0.117 | 22 |
| GEO/02 | 283.250 | 1.000 | 100 | 0.416 | 34 | 0.416 | 92 |
| GEO/05 | 558.500 | 0.466 | 87 | 1.000 | 100 | 0.466 | 95 |
| GEO/07 | 398.500 | 0.000 | 0 | 0.000 | 0 | 0.000 | 0 |
| GEO/08 | 398.500 | 0.358 | 50 | 0.355 | 21 | 0.127 | 25 |
| INF/01 | 7,453.400 | 0.461 | 62 | 0.822 | 78 | 0.379 | 66 |
| ING-IND/25 | 1,195.500 | 1.000 | 100 | 0.418 | 23 | 0.418 | 77 |
| ING-INF/05 | 558.500 | 0.318 | 54 | 0.575 | 21 | 0.183 | 47 |
| MAT/02 | 1,240.250 | 0.180 | 50 | 0.588 | 74 | 0.106 | 55 |
| MAT/08 | 398.500 | 0.541 | 77 | 0.130 | 9 | 0.070 | 28 |
| MAT/09 | 1,572.150 | 0.672 | 71 | 0.346 | 3 | 0.232 | 59 |
| *Total/average* | 45,846.100 | 0.352 | 51 | 0.684 | 55 | 0.228 | 52 |

*Table 5: DEA cost efficiency and percentile (R%) scores for the 23 SDSs at Univ_X*
*\* Percentile rank 0 – 100, with 100 as best and 0 as worst*

## 5. Conclusions

Use of data envelopment analysis for efficiency evaluation of national university and research systems is rapidly taking hold. This tendency is fed by the currently observed tendency towards bibliometrics in evaluation and the generally greater availability of data on inputs and outputs of university systems. However the literature on the issue shows that the knowledge frontier is far from stabilized, and that there is ample room for improvement on various fronts. In this work we have attempted to overcome several critical methodological problems inherent in many of the studies frequently cited as "state of the art". In fact, the potential of DEA in terms of its capacity to depict efficiency in multi-output, multi-input processes has led to a polarization of scholarly attention towards continuous improvement of models through progressive addition of variables, both on the output side and on the side of production factors that determine output. The problem of lack of uniformity in the DMUs observed has not been satisfactorily addressed (as seen in cases where studies resort to a simplistic subdivision between "medical" and "non-medical" universities), or has led to a fragmentation of the analysis and to its application in case studies concerning specific disciplinary areas, in which the DMUs evaluated (usually departments) are still unrealistically considered as uniform. In other words, the literature does not reveal any appreciable attempt to overcome the objective problem of measuring efficiency without field distortions, for purposes of comparative evaluation on a large scale and with a level of detail sufficient to provide robust support for interventions by policy-makers and research institution administrations.

This work opened with the ambitious objective of dealing with these shortcomings and problems, proposing efficiency measurement of the entire Italian university system (78 universities), for each of the 130 scientific disciplinary sectors considered. The proposed level of detail resolves the distorting effects of absence of field-standardization, typical of aggregate analyses. The study offers a level of detail that is highly opportune for those wishing to stimulate better efficiency, because it provides information from highly focused measurement.

The extreme breadth of the field of observation imposes some simplifying hypotheses in the formulation of the DEA inputs and outputs but the study shows that these assumptions are entirely justified in the light of the specific characteristics of the Italian university system. There are clear opportunities to proceed to further refinements in the analysis, and align it to the policy objectives of the governments, for example by considering i) for output, additional indicators such as highly cited papers, concentration of productive scientists, etc.; or indicators for the university activities other than research (didactic activity and technology transfer); ii) for input, the integration of other production factors, such as capital or other types of research staff (PhD students, post-doctoral researchers and fellows, technical personnel, etc.). It would also be interesting to extend the analysis using historic data panels, going beyond static performance to detect the temporal variations resulting from efforts for betterment, for which national assessments should offer appropriate incentive.